# Domain periodicity in an easy-plane antiferromagnet with Dzyaloshinskii-Moriya interaction

Riccardo Tomasello[1*], Luis Sanchez-Tejerina[2], Victor Lopez-Dominguez[3], Francesca Garescì,[4] Anna Giordano,[5] Mario Carpentieri[2], Pedram Khalili Amiri[3], Giovanni Finocchio[5*]

[1]Institute of Applied and Computational Mathematics, FORTH, GR-70013, Heraklion-Crete, Greece

[2]Department of Electrical and Information Engineering, Politecnico of Bari, via Orabona 4, 70125 Bari, Italy

[3]Department of Electrical and Computer Engineering, Northwestern University, Evanston, Illinois 60208, USA

[4]Department of Engineering, University of Messina, I-98166, Messina, Italy

[5]Department of Mathematical and Computer Sciences, Physical Sciences and Earth Sciences, University of Messina, I-98166, Messina, Italy

[*]corresponding authors: gfinocchio@unime.it, rtomasello@iacm.forth.gr

## Abstract

Antiferromagnetic spintronics is a promising emerging paradigm to develop high-performance computing and communications devices. From a theoretical point of view, it is important to implement simulation tools that can support a data-driven development of materials having specific properties for particular applications. Here, we present a study focusing on antiferromagnetic materials having an easy-plane anisotropy and interfacial Dzyaloshinskii-Moriya interaction (IDMI). An analytical theory is developed and benchmarked against full numerical micromagnetic simulations, describing the main properties of the ground state in antiferromagnets and how it is possible to estimate the IDMI from experimental measurements. The effect of the IDMI on the electrical switching dynamics of the antiferromagnetic element is also analyzed. Our theoretical results can be used for the design of multi-terminal heavy metal/antiferromagnet memory devices.

## I. INTRODUCTION

Antiferromagnets (AFMs) are attracting a growing and renewed interest because of the demonstration of their electrical manipulation by spin-orbit torque (SOT), and unique characteristics such as, ultrahigh velocity of domain walls[1–3] and skyrmions[4–7], zero net magnetization[8,9], as well as picosecond switching[10,11] and terahertz dynamics[12,13]. These features pave the way for a number of potential applications in spintronics, ranging from memory and neuromorphic computing devices, to terahertz oscillators[12,13] and detectors[14].

Experimental imaging of the antiferromagnetic order, such as X-ray dichroism, has pointed out the existence of very complex domain patters[15–18], including vortex and antivortex configurations[19,20]. An extended explanation for the pattern structure is attributed to the magnetoelastic energy originating from the substrate that can be strongly spatially non-uniform. However, a tilt of the antiferromagnetic order can be induced by the Dzyaloshinskii-Moriya interaction (DMI) also in ideal systems[21,22] and in the absence of magnetoelastic contributions. The most common devices have an adjacent heavy metal (HM) with large spin orbit coupling, such as Platinum (Pt) interfaced directly with the AFM. In this configuration, we expect the interfacial DMI (IDMI) to play a significant role. Specifically, a systematic study to understand the effect of IDMI on the ground state and dynamics of an AFM has remained elusive to date. Previous results[21] showed that a particular class of materials (hematite $\alpha$-$Fe_2O_3$, iron borate $FeBO_3$, and orthoferrites) characterized by easy-plane anisotropy (EPA) and IDMI exhibit a small net magnetization, due to a small tilting of the spin sublattice originating from the IDMI. Therefore, the corresponding non-zero dipolar field favors the formation of vortices[21].

In this work, we perform micromagnetic simulations showing how the IDMI affects the equilibrium configuration of the Néel vector in collinear (no net magnetization) AFM materials having easy-plane anisotropy. The main result is that the energy contribution originating from a large enough IDMI promotes a non-collinear magnetization orientation[23] thus inducing a ground state characterized by a periodic structure of up and down domains separated by chiral Néel domain walls (NDWs). More interestingly, the periodicity of the domains is strictly connected to the IDMI parameter and can be potentially used for its quantification in AFMs. To this aim, we have derived a simple analytical formula which shows a good agreement with the numerical results achieved within a full micromagnetic framework. Our approach extends to AFMs a method previously developed for ferromagnets to estimate the IDMI constant, which is based on the domain wall size estimation[24]. Our results can be crucial for developing an approach to estimate the IDMI in AFMs, also because other standard procedures developed for ferromagnets, such as Brillouin light scattering (BLS)[25–28] and asymmetric expansion of a bubble domain[24,29], cannot be directly applied to AFMs. We further show the implications of the presence of the periodic domain structures in the design of multi-terminal antiferromagnetic memory devices. The paper is organized as follows. Section II describes the device geometry and parameters as well as the micromagnetic model. Section III deals with the development of the analytical theory to estimate the NDW periodicity. Section IV shows the results regarding the ground state of the magnetizations together with a comparison between the analytical theory and micromagnetic model periodicities. Section V presents the dynamics of both NDWs and uniform state driven by an in-plane electrical current, which can be used to design antiferromagnetic memory device and Section VI summarizes the conclusions.

## II. DEVICE STRUCTURE AND MICROMAGNETIC MODEL

We investigate a circular AFM pillar built on top of a HM underlayer (Pt), in a 4-terminal device, as shown in Figure 1. The AFM has a 400 nm diameter and a 6 nm thickness. In Fig. 1(a), a Cartesian coordinate system is also introduced, with the $z$-axis being the out-of-plane direction, and the $x$ and $y$-axes the in-plane directions. Figure 1(b) shows the spatial distribution of the current density flowing in the Pt heavy metal and the AFM (inset), as computed by finite element simulations[30] when the current is applied between the A-A' terminals. We observe that the AFM diameter has to be smaller than half of the HM width in order to obtain a uniform current distribution in the AFM (see green circle and corresponding current distribution). If we consider a HM width of 1000 nm, we can fix the AFM diameter at 400 nm in this study.

The micromagnetic calculations are based on a continuous model which describes the antiferromagnetic order by considering two sublattices characterized by a normalized magnetization vectors $\mathbf{m}_1 = \mathbf{M}_1 / M_s$ and $\mathbf{m}_2 = \mathbf{M}_2 / M_s$, respectively ( $M_s$ is the saturation magnetization of the two sublattices $M_{s1} = M_{s2} = M_s$ ). The AFM static properties are studied numerically by solving two coupled LLG equations[3,13]

$$\begin{cases} \dfrac{d\mathbf{m}_1}{dt} = -\gamma_0 \mathbf{m}_1 \times \mathbf{H}_{\text{eff},1} + \alpha \mathbf{m}_1 \times \dfrac{d\mathbf{m}_1}{dt} \\ \dfrac{d\mathbf{m}_2}{dt} = -\gamma_0 \mathbf{m}_2 \times \mathbf{H}_{\text{eff},2} + \alpha \mathbf{m}_2 \times \dfrac{d\mathbf{m}_2}{dt} \end{cases} , \qquad (1)$$

where $\gamma_0$ is the gyromagnetic ratio, $\alpha$ is the Gilbert damping parameter, and $\mathbf{H}_{\text{eff},1}$ and $\mathbf{H}_{\text{eff},2}$ are the effective fields for the first and second sublattice, respectively. Both effective fields include the exchange, easy-plane anisotropy, as well as the IDMI contributions. The total energy density can be written as

$$\varepsilon_{\text{tot}} = \varepsilon_{\text{exch}} + \varepsilon_{\text{ani}} + \varepsilon_{\text{IDMI}} \qquad , \qquad (2)$$

where

$$\varepsilon_{\text{exch}} = A_{11}\left(\nabla \mathbf{m}_1\right)^2 + A_{11}\left(\nabla \mathbf{m}_2\right)^2 + A_{12}\left(\nabla \mathbf{m}_1\right)\left(\nabla \mathbf{m}_2\right) + \frac{4A_0}{a^2}\mathbf{m}_1\mathbf{m}_2$$

$$\varepsilon_{\text{ani}} = K_u\left(1 - \left(\mathbf{m}_1\mathbf{u}_z\right)^2\right) + K_u\left(1 - \left(\mathbf{m}_2\mathbf{u}_z\right)^2\right) \tag{3}$$

$$\varepsilon_{\text{IDMI}} = D\left[\left(\mathbf{m}_1 \cdot \mathbf{u}_z\right)\vec{\nabla}\cdot\mathbf{m}_1 - \mathbf{m}_1 \cdot \vec{\nabla}\left(\mathbf{m}_1 \cdot \mathbf{u}_z\right)\right] + D\left[\left(\mathbf{m}_2 \cdot \mathbf{u}_z\right)\vec{\nabla}\cdot\mathbf{m}_2 - \mathbf{m}_2 \cdot \vec{\nabla}\left(\mathbf{m}_2 \cdot \mathbf{u}_z\right)\right]$$

being $\mathbf{u}_z$ the unit vector along the out-of-plane direction. From Eq. (3), one can derive each term of the two effective fields. In particular, the exchange fields include three contributions:

$$\mathbf{H}_{1,\text{exch}} = \frac{2A_{11}}{\mu_0 M_s}\nabla^2\mathbf{m}_1 + \frac{4A_0}{a^2\mu_0 M_s}\mathbf{m}_2 + \frac{A_{12}}{\mu_0 M_s}\nabla^2\mathbf{m}_2,$$

$$\boldsymbol{H}_{2,\text{exch}} = \frac{2A_{11}}{\mu_0 M_s}\nabla^2\mathbf{m}_2 + \frac{4A_0}{a^2\mu_0 M_s}\mathbf{m}_1 + \frac{A_{12}}{\mu_0 M_s}\nabla^2\mathbf{m}_1. \tag{4}$$

where $\mu_0$ is the vacuum permeability, and $a$ is the lattice constant. In Eqs. (3) and (4), $A_{11} > 0$ is the inhomogeneous intra-lattice contribution, $A_{12} < 0$ is the inhomogeneous inter-sublattice contribution, and $A_0 < 0$, is the homogeneous inter-sublattice contribution to the exchange energy. The expressions for the IDMI fields are

$$H_{\text{IDMI},1} = -\frac{2D}{\mu_0 M_S}\left(\mathbf{u}_z\left(\nabla \cdot \mathbf{m}_1\right) - \nabla m_{1,z}\right),$$

$$H_{\text{IDMI},2} = -\frac{2D}{\mu_0 M_S}\left(\mathbf{u}_z\left(\nabla \cdot \mathbf{m}_2\right) - \nabla m_{2,z}\right), \tag{5}$$

where $D$ is the IDMI parameter, and $m_{1,z}$ and $m_{2,z}$ are the out-of-plane components of the magnetization of the first and second sublattice, respectively. Additionally, the IDMI also affects the boundary conditions by imposing a field $\mathbf{H}_{\text{IDMI},iS} = \frac{D}{\mu_0 M_S}\left(\mathbf{m}_i \times \left(\mathbf{n} \times \mathbf{u}_z\right)\right)$ at the lateral edges ($x$ and $y$ axes) of the sample, where $i = 1, 2$. Therefore, the boundary conditions for the i-th sublattice are modified[3] as

$$2A_{11}\partial_n\mathbf{m}_i + A_{12}\mathbf{m}_i \times \left(\partial_{\mathbf{n}}\mathbf{m}_j \times \mathbf{m}_i\right) + D\mathbf{m}_i \times \left(\mathbf{n} \times \mathbf{u}_z\right) = \mathbf{0}, \tag{6}$$

where $j = 1, 2; \ j \neq i$. The anisotropy fields are

$$H_{\text{ani},1} = \frac{2K_u}{\mu_0 M_S}\mathbf{m}_1\mathbf{u}_\text{k},$$

$$H_{\text{ani},2} = \frac{2K_u}{\mu_0 M_S}\mathbf{m}_2\mathbf{u}_\text{k}. \tag{7}$$

with $K_u$ being the anisotropy constant. We used a 4 x 4 x 6 nm³ discretization cell, and fixed $M_s =$ 400 kA/m, and $A_0 = -0.5$ pJ/m. The static results do not change in the range $300 \le M_s \le 500$ kA/m and $-20 \le A_0 \le -5$ pJ/m (see Note 1 in the Supplemental Material).

## III.    ANALYTICAL THEORY

The Euler-Lagrange equations for a point inside the sample, considering the energy given in Eq. (3), are

$$\left\{ -2A_{11}\left(\nabla^2 \mathbf{m}_i\right) - A_{12}\left(\nabla^2 \mathbf{m}_j\right) + \frac{4A_0}{a^2}\mathbf{m}_j - 2K_u\left(\mathbf{m}_i \cdot \mathbf{u}_z\right)\mathbf{u}_z - \right.$$
$$\left. -2D\left[\nabla\left(\mathbf{m}_i \cdot \mathbf{u}_z\right) - \left(\nabla \cdot \mathbf{m}_i\right)\mathbf{u}_z\right]\right\} \times \mathbf{m}_i = 0 \tag{8}$$

By considering the following hypotheses: (*i*) the modulus of the sublattice magnetization is constant, $|\mathbf{m}_i| = 1$, (*ii*) the two sublattice magnetizations are perfectly aligned antiparallel to each other, i.e. $\mathbf{m}_1 = -\mathbf{m}_2 \rightarrow \nabla^2 \mathbf{m}_1 = -\nabla^2 \mathbf{m}_2$, which is true at equilibrium, and (*iii*) the rotation of the magnetization takes place in a fixed plane, we can write for each sublattice that

$$\frac{\partial^2 \theta_i}{\partial x^2} = -\frac{|K_u|\sin\theta_i\cos\theta_i}{A_{11} - A_{12}/2}$$
$$\frac{\partial \varphi_i}{\partial x} = 0 \tag{9}$$

where $\theta_i$ is the angle of rotation with respect to an arbitrary axis lying in the plane, and $\varphi_i$ is the angle of rotation with respect to the plane, which is assumed to be constant and equal to zero (assumption (*iii*)). Since the same equation is valid for both sublattices, we will omit subindeces without losing generality. Notice that Eq. (9) is formally the same as in the case of ferromagnets[31], where the exchange parameter $A$ has been replaced by the effective exchange $2A_{11} - A_{12}$, so we can straightforwardly apply the same procedure already developed for ferromagnets.

First, we consider the special case of isotropic media, that is $K_u = 0$. Therefore, Eq. (9) becomes $\frac{\partial^2 \theta}{\partial x^2} = 0$ and thus $\frac{\partial \theta(x)}{\partial x} = \frac{2\pi}{\lambda_0}$, where $\lambda_0$ is a constant of integration in units of *meter*, giving the periodicity. Inserting this condition in the energy density of Eq. (3) and minimizing the energy with respect to $\lambda_0$ we obtain the periodicity

$$\lambda_0 = 2\pi \frac{\left(2A_{11} - A_{12}\right)}{D} = 2\pi\xi, \tag{10}$$

which is a function of the ratio between the IDMI and the (inhomogeneous) exchange. In the case $K_u \neq 0$, Eq. (9) yields to

$$\frac{d\theta}{\sqrt{C - \sin^2 \theta}} = \frac{dx}{\Delta} \ , \tag{11}$$

where $C$ is an integration constant and $\Delta = \sqrt{(2A_{11} - A_{12})/(2K_u)}$ is the static domain wall width for AFM[3]. Integrating over a quarter of a period, it gives a periodicity $\lambda$

$$\lambda = 4\Delta \int_0^{\pi/2} \frac{d\theta}{\sqrt{C - \sin^2 \theta}} \ , \tag{12}$$

which depends on the first-kind elliptic integral. In order to determine the integration constant $C$, we minimize the energy density with respect to the periodicity $\lambda$

$$\frac{D}{D_c} = \frac{\pi^2 \Delta}{\lambda_0} = \frac{\pi \Delta}{2\xi} = \int_0^{\pi/2} \sqrt{C - \sin^2 \theta} \ , \tag{13}$$

where $D_c = \frac{2}{\pi} \sqrt{(2A_{11} - A_{12}) 2|K_u|}$ is the minimum IDMI needed to get the cycloid state, and the right-hand term is the second-kind elliptic integral.

## IV. RESULTS

### A. Statics

Figure 2 summarizes the snapshots of the ground states of the circular AFM as a function of the EPA constant ($K_u$) and the IDMI parameter ($D$) - the colormap codes the out-of-plane component of the sublattice 1, which also coincides with the one of the Néel vector. The ground state at low IDMI corresponds to the uniform configuration of the Néel vector, while a larger IDMI energy fosters the formation of out-of-plane domains separated by NDWs (the in-plane component of the magnetization within the domain wall is perpendicular to the direction of the domain wall), which can be oriented in each direction inside the x-y plane due to the EPA. We can consider two scenarios characterized by zero and non-zero EPA, respectively. In the former, the out-of-plane domains result from the competition between only the exchange and IDMI energies. The reason is that, while the exchange promotes the parallel alignment of the magnetization, the IDMI promotes a misalignment, which gradually tilts the local spins in the same direction of rotation (i.e. the IDMI creates a chiral effect). This rotation takes place in the plane formed by the vector perpendicular to the interface, and the vector linking both spatial positions. Consequently, out-of-plane domains separated by NDWs are created, and the DW periodicity is obtained as the ratio between the exchange and the IDMI energies, which determines the deviation angle. A more complex situation occurs when the space is not isotropic, which corresponds to the non-zero EPA case. In that case, a deviation from the circular path towards an ellipse takes place because the rotation of the spin is slower (or even null) when the anisotropy stabilizes the orientation (aphelion) and faster when it destabilizes the orientation (perihelion). A systematic study based on micromagnetic simulations confirms that $M_s$ and $A_0$ do not affect the results, while $A_{11}$ (see Note 1 in the Supplemental Material) and $A_{12}$ change the periodicity, as shown in the next paragraph.

## B. A comparison between numerical and analytical calculations

Figure 3 displays a comparison between the micromagnetic and analytical periodicity for different values of the IDMI parameter and EPA constant. In the micromagnetic simulations, the periodicity is computed as the distance between two consecutive identical magnetization values (see inset of Fig. 3(a)), while it is analytically calculated by using Eq. (10) for zero EPA and Eqs (12) and (13) for finite EPA values. We wish to highlight that to calculate numerically the periodicity at low $K_u$ with a better resolution, we have simulated larger cross section (not shown). Figure 3(a) shows the periodicity dependence on IDMI constant at zero $K_u$ as a function of the inhomogeneous inter-sublattice exchange constant $A_{12}$. For each value of $A_{12}$, the analytical period decreases with increasing $D$, confirming that the IDMI promotes the proliferation of NDWs, as also obtained by micromagnetic simulations (see also Fig. 2). On the other hand, for a fixed $D$ value, the period is larger as the magnitude of $A_{12}$ increases. This feature points out that in the continuous model of AFMs, as the micromagnetic one used in this work, the role of the inhomogeneous inter-sublattice exchange term is non-negligible and should be considered for the correct understanding of the AFM ground state. Figure 3(a) also shows an excellent agreement between the micromagnetic and analytical results.

Figure 3(b) displays the periodicity dependence on $A_{12}$ for three values of $D$ and for a non-zero $K_u$=-0.10 x $10^5$ J/m$^3$. Similar conclusions can be drawn, i.e. the periodicity increases with $A_{12}$ and decreases with $D$. Again, the analytical outcomes fit well with the micromagnetic ones. However, we wish to underline that for the point $D = 0.40$ mJ/m$^2$, $A_{12}$= 8 pJ/m, vortex cores are stabilized (see snapshot in the Supplementary Note 2). This means that our analytical theory does not apply for that point, despite the good match.

## C. IDMI parameter estimation

Our model extends the method previously developed for IDMI estimation in ferromagnets, based on the domain size calculation[24], and represents a possible tool to estimate the IDMI and exchange parameters from experimental images of NDW patterns in AFMs. This is important since other methods for IDMI measurement, such as spin wave nonreciprocity measurement via Brillouin light scattering (BLS)[25–28], and asymmetric expansion of a bubble domain[24,29], both of which are used in ferromagnets, cannot be similarly used in AFM materials.

Our approach can be applied through the following steps. From the experimental measurements on bulk materials, we can estimate the value of $|K_u|$, the NDW periodicity $\lambda$, and the NDW width $\Delta = \sqrt{\dfrac{2A_{11} - A_{12}}{2|K_u|}}$. From the two latter values, we can calculate the value of the elliptic integral of the first kind (Eq. (12)), and so the value of the argument of this function. Therefore, combining Eq. (12) and Eq. (13) give us the value of the IDMI

$$D = \frac{16|K_u|\Delta^2}{\pi\lambda} E\left(\frac{1}{\sqrt{C}}\right) K\left(\frac{1}{\sqrt{C}}\right) \tag{14}$$

where $K\left(\dfrac{1}{\sqrt{C}}\right)$ and $E\left(\dfrac{1}{\sqrt{C}}\right)$ are the first and second kind elliptic integrals.

# V. APPLICATION AS AN ANTIFERROMAGNETIC MEMORY DEVICE

## A. Background

The AFM order can be manipulated by using exchange bias[32–34], strain[35,36], femtosecond lasers[37] and electrical currents. As well-established for ferromagnets[38–42], the current-induced manipulation of AFMs is very promising because it allows spintronic memories to be implemented alongside transistors in electronic circuits[43], with electrical read and write operations. From a fundamental point of view, the electrical switching of AFM relies on the local transfer of spin-angular momentum to the alternating spins, which then promotes a rigid rotation of the whole lattice in a different direction. In a continuous formulation of this phenomenon, the Néel vector switches from one direction to the other one depending on the spin-polarization of the applied electric current. The Néel vector can be read out via the anisotropic and spin-Hall magnetoresistance effects, and depending on its orientation it can be used as a binary memory (coding the bits '0' and '1'), or a memristive system (analog memory coding multiple states) when the ground state can have multiple domains. A typical geometry designed for AFM switching is a multi-terminal device, which enables the writing operation through current pulses applied along different device terminals, and the readout via either the transversal resistivity (anomalous Hall, anisotropic, spin-Hall resistance) or the longitudinal one (planar Hall effect). Wadley et al.[15] and Bodnar et al.[44] observed AFM switching in CuMnAs and $Mn_2Au$, respectively, by applying a number of consecutive current pulses and using the AMR as a readout mechanism. The switching process occurred via domain wall reorientation. Similar results were achieved by Grzybowski et al.[16] but they observed local switching in regions of 100-200 nm in size, hence they ascribed this to the magnetoelastic deformation. A different system has been proposed by Moriyama et al.[17], who designed a Pt/NiO/Pt 4-terminal device and electrically detected the two AFM order states by spin-Hall magnetoresistance. However, these previous works relied on materials which are hard to be integrated in conventional semiconductor memory manufacturing technology[9,15,16,43,45,46]. Recently, Shi et al.[47] demonstrated switching dynamics in PtMn in contact with a Pt or Ta heavy metal, which are standard materials used in existing magnetic tunnel junctions, and therefore easily integrable with state-of-the-art silicon technology[48,49]. For this reason, our theoretical study is based on PtMn magnetic material parameters.

## B. Micromagnetic model

In order to study the AFM order dynamics, we add the following torques[47] to Eq. (1):

$$\begin{cases} \mathbf{T}_1 = d_J \left( \dfrac{\theta_{i-DLT} J_{HM}}{t_{AFM}} + \theta_{b-DLT} J_{AFM} \right)(\mathbf{m}_1 \times \mathbf{m}_1 \times \mathbf{p}) + d_J P J_{AFM} \nabla \mathbf{m}_1 - d_J P J_{AFM} \beta \mathbf{m}_1 \times \nabla \mathbf{m}_1 \\ \mathbf{T}_2 = d_J \left( \dfrac{\theta_{i-DLT} J_{HM}}{t_{AFM}} + \theta_{b-DLT} J_{AFM} \right)(\mathbf{m}_2 \times \mathbf{m}_2 \times \mathbf{p}) + d_J P J_{AFM} \nabla \mathbf{m}_2 - d_J P J_{AFM} \beta \mathbf{m}_2 \times \nabla \mathbf{m}_2 \end{cases} \quad (15)$$

where $d_J$ is a torque coefficient given by $d_J = \dfrac{g \mu_B}{2 e M_S^2}$, where $g$ is the Landè factor, $\mu_B$ is the Bohr magneton, and $e$ is the electron charge. The first term of both Eqs. (15) represents the sum of the interfacial-damping-like torque (IDLT)[2] and bulk-damping-like torque (BDLT)[1]. The coefficient $\theta_{i-DLT}$ takes into account the efficiency of the charge/spin current conversion of the current $J_{HM}$ flowing in the HM due to mechanisms like spin-Hall and spin-galvanic effects. As the thickness of the AFM, $t_{AFM}$ increases, this effect is reduced proportionally. On the other hand, $\theta_{b-DLT}$ describes the efficiency of the relativistic spin–orbit coupling in generating spin-current from the charge current $J_{AFM}$ flowing through the metallic AFM. This latter mechanism, originating directly in the bulk, does not depend on the $t_{AFM}$. In our model, we consider $J_{HM} = J_{AFM}$ and $(\theta_{i-DLT} + t_{AFM} \theta_{b-DLT})$=0.15[47]. The vector $\mathbf{p}$ is the direction of the spin- polarization (y-direction for a voltage applied across A-A', see Fig. 1(a)). The latter two terms of Eqs. (15) represent the Zhang-Li spin-transfer torques (STT)[50] originating from the antiferromagnetic textures, composed of adiabatic and non-adiabatic contributions directly proportional to the current flowing in the AFM. $P$=0.7 is the spin-polarization, and $\beta = 0.05$ is the non-adiabatic term.

## C. Results

In the following, we compute the switching time - current relations for 4 values of the IDMI parameter, when the EPA constant is fixed to -0.10 x $10^5$ J/m$^3$ and the electrical current is applied along the x-direction (terminals A-A'). We define the switching time as the time interval until the y-component of the Néel vector reaches the 95% its final value. We first describe the results when only the IDLT and BDLT act on the AFM (i.e. no STT). For $D$=0.00 and 0.20 mJ/m$^2$, the ground state is uniform in the x-direction, while for $D$=0.60 and 0.80 mJ/m$^2$, we obtain out-of-plane domains (as previously shown in Fig. 2). For the latter cases, we first applied a sufficiently large current density > 10 MA/cm$^2$ in order to orient all the random initial NDWs along the x-direction. Analogous results are achieved if the electrical current is applied along the y-direction (terminals B-B') and the initial in-plane Néel vector is aligned along the y-direction.

We plot, in Fig. 4(a), the switching results where we only report switching time smaller than 20 ns. Regardless of the ground state, the switching mechanism is characterized by a 90° rotation of the in-plane component of the Néel vector towards the direction of the spin-polarization. In particular, for small current densities $\leq$ 7.0 MA/cm$^2$, the NDWs switch faster than the uniform state, whereas for $J_{\mathrm{HM}} > 7$ MA/cm$^2$, the switching time is nearly the same for all the cases. As the IDMI increases, there a qualitative change in the switching mechanism, at low IDMI there is a uniform domain rotation while as the ground state becomes non-uniform, the switching is due to a domain rearrangement. As expected for an EPA AFM, the domain rotation is mainly driven by the IDLT and BDLT, which act as an effective out-of-plane field $\mathbf{H}_{DLT,i} \propto J_{HM}\left(\mathbf{m}_i \times \mathbf{p}\right)$, thus reorienting the in-plane Néel vector along the $y$-axis (see MOVIE 1 for $D$=0.20 mJ/m$^2$, $J_{\mathrm{HM}}$=10 MA/cm$^2$ and $\mathbf{p} \equiv \mathbf{u}_\mathrm{y}$). In the case of domain rearrangement, the switching is dominated by their motion and the final alignment of the DW along the direction of the spin-polarization (as in the case of uniform rotation) to minimize the energy. In detail, the initial NDWs are shifted perpendicularly to the spin-polarization direction, as it occurs for the 1D SHE-driven NDW motion[1–3], and, subsequently, more NDWs are nucleated from the sample edges. The switching finishes once all the initial perpendicular-to-the-spin-polarization NDWs are expelled from the system and replaced by horizontal NDWs parallel to the spin-polarization direction (see MOVIE 2 for $D$=0.60 mJ/m$^2$, $J_{\mathrm{HM}}$=10 MA/cm$^2$ and $\mathbf{p} \equiv \mathbf{u}_\mathrm{x}$). We wish to highlight one more time that at low current the domain rearrangement is faster than the uniform rotation because the large velocity of the domain wall motion[1,3] induced by the SOT as compared to the uniform rotation driven by the change in the field gradient originated by the SOT.

Next, we also consider the STT. It has no effect on the uniform ground state ( $\nabla \mathbf{m}_1 = \nabla \mathbf{m}_2 = 0$ ), while it promotes a NDWs translation along the electrical current direction[47]. However, for the range of currents considered here, these shifting dynamics are negligible compared to the 90° rotation induced by the IDLT and BDLT linked to the SHE.

The above described SHE-switching dynamics can be exploited in the four-terminal device depicted in Fig. 1(a) to design AFM memories. The information is coded in the direction of the in-plane Néel vector which rotates 90° during the switching process. We define the digital bits '0' and '1' as being represented by the Néel vector along the x- and y-direction, respectively. The writing protocol starts with the application of a sufficiently large initialization current between the terminals B-B', in order to orient the initial random NDWs in the same direction (x-direction, bit '0'). If the other digital bit needs to be written, the current is applied between the terminals A-A'. The reading process occurs via the same terminals, e.g. B-B', where the signal derived from the in-plane component of the Néel vector is detected. It is noteworthy that in this device concept, a single-domain AFM is not required in order to allow the device to work as a memory device with electrical readout. This is because the presence of NDWs due to the IDMI ensures that the in-plane component of the Néel vector is fully aligned along either the x or y axis in all of the domain walls in the '0' and '1' states, thus allowing for distinction between the two states when reading out using an electrical readout method such as AMR.

## VI.    SUMMARY AND CONCLUSIONS

In summary, we have micromagnetically shown that a sufficiently large IDMI promotes the formation of periodic domain patterns as ground state of an AFM characterized by an EPA. The periodicity of those domain patterns can be calculated by an analytical model. This allows us to extend to AFMs the well-known approach used in ferromagnets for estimation of the DMI value. The analytical periodicity is useful to estimate the IDMI parameter in AFMs, once the anisotropy constant is known. We further showed that a spin-polarized current can orient both the uniform and NDW states along the direction of the spin-polarization. Such switching dynamics can be exploited in a four-terminal device to implement AFM memories based on a 90° reorientation of the Néel vector.

ACKNOWLEDGEMENT

This work has been supported by the project "ThunderSKY" funded by the Hellenic Foundation for Research and Innovation and the General Secretariat for Research and Technology, under Grant No.

871. This work was partially supported by the PETASPIN association. This work was also in part supported by a grant from the U.S. National Science Foundation, Division of Electrical, Communications and Cyber Systems (NSF ECCS-1853879).

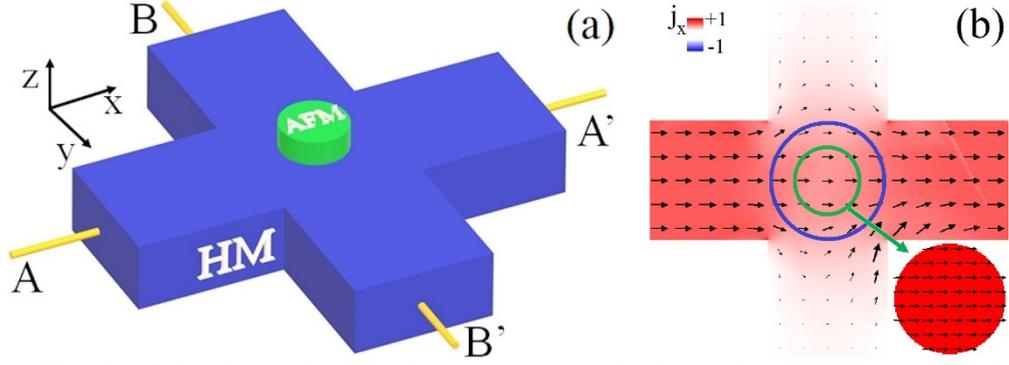

Figure 1: (a) Sketch of the 4-terminal device structure under investigation along with the Cartesian coordinate system. (b) Spatial distribution of the electrical current density through the Pt heavy metal and the AFM (inset). The green circle represents the circular AFM under investigation (400 nm in diameter), where the current distribution is uniform, while the blue circle represents a larger AFM, where the current distribution is non-uniform. The colors are linked to the *x*-component of the current density, as indicated in the bar.

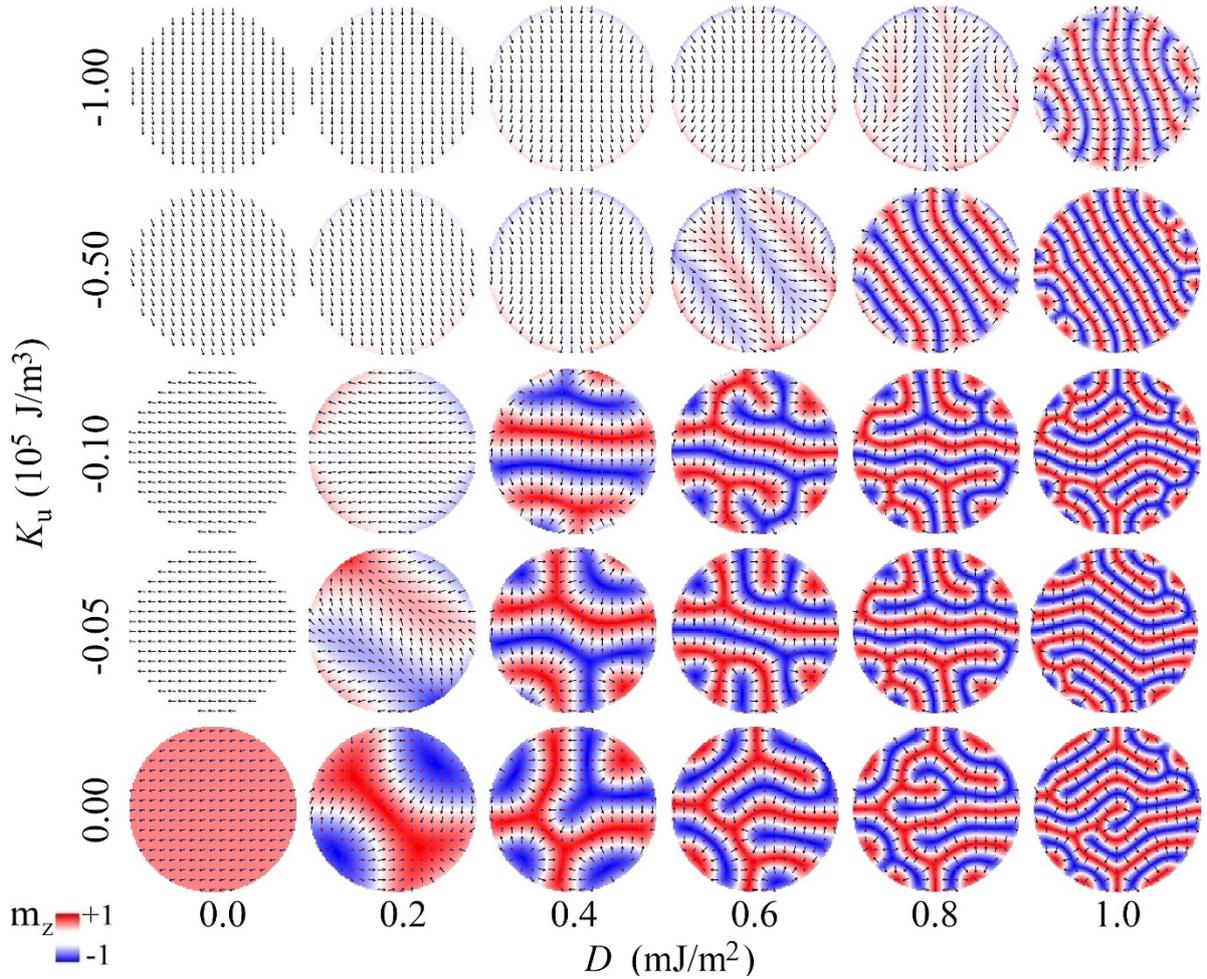

Figure 2: Snapshots of the ground states of the magnetization of sublattice 1 for different combinations of the IDMI and EPA parameters.

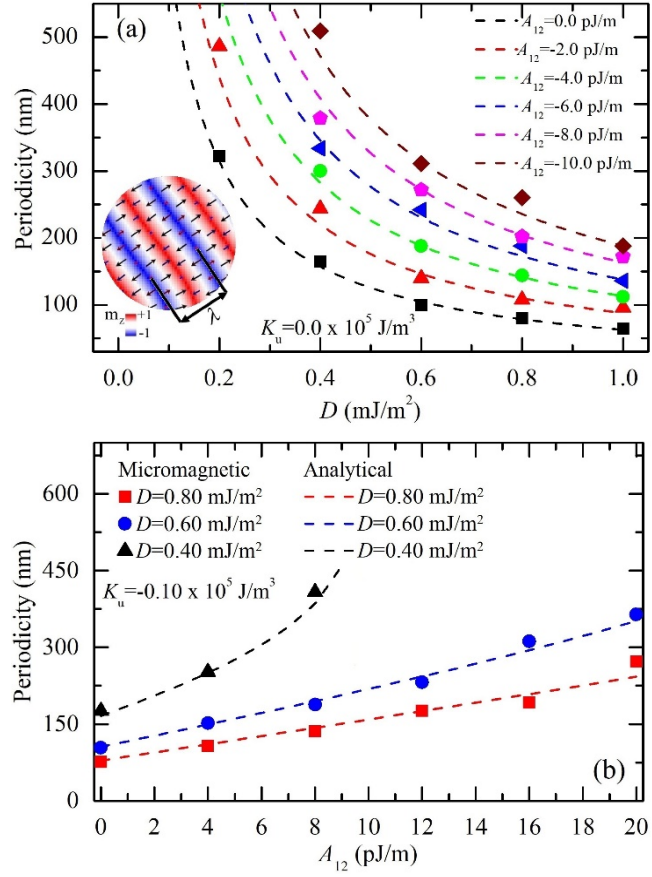

Figure 3: A comparison of the micromagnetic (symbols) and analytical (dashed lines) domains periodicity (a) as a function of the IDMI, for different values of $A_{12}$ at zero $K_u$, and (b) as a function of $A_{12}$ for three values of $D$ and for $K_u$ = -0.10 x $10^5$ J/m$^3$. The analytical results are calculated using (a) Eq. (10) and (b) Eqs. (12) and (13). The inset in (a) shows a magnification of a snapshot where the micromagnetic period is indicated. The colors represent the z-component of the magnetization of the sublattice 1, as indicated in the color bar.

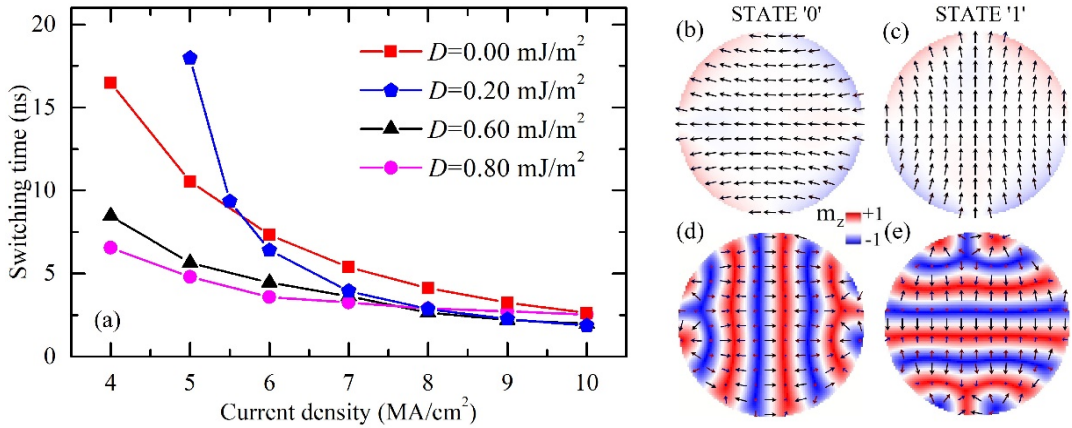

Figure 4: (a) switching time as a function of the current density for different values of the IDMI parameter, therefore of the AFM ground state (Uniform or DW), and for $K_u$=-0.10 x $10^5$ J/m$^3$. (b)-(e) spatial distribution of the 1$^{st}$-sublattice magnetization corresponding to the initial (state '0') and final (state '1') configurations when $D$=0.20 mJ/m$^2$ ((b) and (c)) and $D$=0.60 mJ/m$^2$ ((d) and (e)).